\documentstyle[abe,ijmp,12pt]{article}
\begin{document}
\input mssymb.tex
%
\pagestyle{empty}
\setlength{\topmargin}{1truecm}   
\baselineskip20pt
\font\hugeit=cmti10 scaled \magstep4
\def\today{\ifcase\month\or
  January\or February\or March\or April\or May\or June\or
  July\or August\or September\or October\or November\or December\fi
  \ \ \number\year}
\hrule height0pt depth0pt
\vspace*{-80pt}
\rightline{\bf RIMS-1208}
\vspace*{40pt}
\centerline{\cmssB Construction of an Identically Nilpotent BRS Charge}
\vskip10pt
\centerline{\cmssB in the Kato-Ogawa String Theory}
\vskip50pt
\centerline{
 Mitsuo Abe\foot(*,{E-mail: \abemail}) 
 }
\centerline{\it Research Institute for Mathematical Sciences,
Kyoto University, Kyoto 606-8502, Japan}
\vskip3pt
\centerline{ and }
\vskip3pt
\centerline{
 Noboru Nakanishi\foot({\dagger},{Professor Emeritus of Kyoto 
 University.  E-mail: \nnmail})}
\centerline{\it 12-20 Asahigaoka-cho, Hirakata 573-0026, Japan}
\vskip20pt
\centerline{ July 1998}

\vskip100pt
\centerline{\bf Abstract}
In previous work, the conformal-gauge two-dimensional quantum gravity
in the BRS formalism has been solved completely in terms of Wightman 
functions.  In the present paper, this result is extended to the closed
and open bosonic strings of finite length; the open-string case is 
nothing but the Kato-Ogawa string theory.  The field-equation anomaly 
found previously, which means a slight violation of a field equation 
at the level of Wightman functions, remains existent also in the 
finite-string cases.  By using this fact, a BRS charge nilpotent even 
for $D\not=26$ is explicitly constructed in the framework of the 
Kato-Ogawa string theory.
The FP-ghost vacuum structure of the Kato-Ogawa theory is made more 
transparent; the appearance of half-integral ghost numbers and the 
artificial introduction of indefinite metric are avoided.

\vfill\eject
\pagestyle{plain}
\setlength{\oddsidemargin}{.5truecm}
\setlength{\textheight}{23.cm}  
\setlength{\textwidth}{16.cm}
\setlength{\topmargin}{-.5cm}
\setlength{\baselineskip}{19.8pt}
\setlength{\parindent}{25pt}
\textfont0=\tenrm  \textfont1=\teni \textfont2=\tensy \textfont3=\tenex
\def\rm{\fam0 \tenrm} \def\mit{\fam1 } \def\cal{\fam2 }
\def\bf{\tenbf}  \def\it{\tenit} \def\sl{\tensl}
\scriptfont0=\sixrm  \scriptfont1=\sixi  \scriptfont2=\sixsy
\scriptscriptfont0=\smallr \scriptscriptfont1=\smalli 
                           \scriptscriptfont2=\smallsy

\rm
%
\Sec{Introduction}
In 1983, Kato and Ogawa\cite{Kato-Ogawa} published a fundamental paper 
on the BRS quantization of the bosonic string of finite length based on 
the Lagrangian density of the conformal-gauge two-dimensional quantum 
gravity together with open-string boundary conditions.  
According to their conclusion, the square of the normal-ordered BRS 
charge written in terms of the creation and annihilation operators of 
string and FP-ghosts is {\it nonvanishing unless $D=26$ and 
$\alpha_0=1$\/}, where $D$ stands for the dimension of the world 
in which the string lives and $\alpha_0$ denotes a regularization 
parameter of the Hamiltonian (interpreted as the zero intercept of 
the leading Regge trajectory).  Then they established no-ghost theorem 
for the string only for $D=26$ and $\alpha_0=1$.
\par
The purpose of the present paper is to point out that Kato-Ogawa's 
conclusion on the BRS charge is not intrinsic; more precisely speaking, 
we show that {\it it is possible to construct explicitly an identically 
nilpotent BRS charge in the Kato-Ogawa framework.}  
One may wonder why two different BRS charges exist in one particular 
theory.  
The key word for answering this question is ``field-equation anomaly''.
So, we first explain what the field-equation anomaly is.
\par
In the Heisenberg picture, field equations and equal-time 
(anti)commutation relations uniquely determine the full-dimensional 
(anti)commutation relations at least in principle.  
This problem can be explicitly worked out in the two-dimensional 
quantum gravity in various gauges such as de Donder gauge,\cite{AN1} 
light-cone gauge,\cite{AN2} conformal gauge,\cite{AN3} etc.\cite{AN4}  
That is, in each of those models, the algebra of field operators is 
completely found in closed form.  
The next problem is to represent this algebra in terms of state vectors 
so as to be consistent with certain physically natural requirements.  
The representation can be explicitly constructed by giving the set of 
all Wightman functions, i.e., vacuum expectation values of 
field-operator products. This representation is, of course, consistent 
with the full-dimensional (anti)commutation relations, but not always 
consistent with field equations owing to the presence of singular 
products in them.  
That is, we encounter a kind of anomaly, which we call ``field-equation 
anomaly''.  The existence of the field-equation anomaly has been found 
in each model stated above.
One should note that the violation of a field equation is very slight 
in the sense that by differentiating it once or twice we can find an 
anomaly-free field equation having the same degrees of freedom as that 
of the original equation.
\par
Now, one can understand the essence of the anomaly problem of the BRS 
charge by the following remarks.
\begin{item1}
\item[1.]  
The expression for the square of the BRS charge given by Kato and 
Ogawa\cite{Kato-Ogawa} {\it cannot be obtained without using the Fock 
representation\/}, though they wrote as if it had been obtained by 
straightforward operator calculation.
\item[2.]
Kato and Ogawa\cite{Kato-Ogawa} eliminated the B field (after its field 
redefinition) by regarding its field equation as an identity, that is, 
in their theory the B field equation is an equality which holds even 
at the representation level.
\item[3.]
In our previous paper,\cite{AN3} we have shown in the BRS formalism of 
the conformal-gauge two-dimensional quantum gravity that the B field 
equation suffers from the field-equation anomaly, which disappears 
if and only if $D=26$.  By using the B field equation at the operator 
level, therefore, it is possible to rewrite the BRS charge into the 
one which is anomaly-free, and hence identically nilpotent, at the 
representation level.
\item[4.] 
As is shown in the present paper, the essential results obtained in our 
previous paper can be transcribed into the case of finite string with 
some minor modifications.
\end{item1}
\par
In the framework of the Kato-Ogawa theory, we rewrite their BRS charge 
by using the original form of the B field equation just as has been 
done in our previous paper.  We then obtain a BRS charge which is 
completely nilpotent even at the representation level.  Thus one can no 
longer claim that the critical dimension $D=26$ is a consequence of the
requirement of the BRS invariance in the Kato-Ogawa theory.  
Rather, we should say the $D=26$ is the condition for the absence of 
the field-equation anomaly in conformal gauge (but it is not so in 
de Donder gauge\cite{AN1}).
\par
As a side remark, we discuss the FP-ghost vacuum structure.  
Although Kato and Ogawa\cite{Kato-Ogawa} artificially introduced 
a rather complicated FP-ghost vacuum structure, we show that it can be 
reformulated into a more natural one.  As a consequence, we can avoid 
the appearance of half-integral ghost numbers and also the introduction 
of the indefinite metric which is inconsistent with the hermiticity of 
the original action.
\par
The present paper is organized as follows.  
In Sec.2, we review the main results of the conformal-gauge 
two-dimensional quantum gravity obtained in our previous paper.  
In Sec.3, we extend them into the case of the closed string of finite 
length.
In Sec.4, we reformulate the Kato-Ogawa open-string theory into the 
formalism similar to ours.  
In Sec.5, the FP-ghost vacuum structure of the Kato-Ogawa theory is 
shown to be made more transparent.
In Sec.6, we explicitly construct an identically nilpotent BRS charge 
in the framework of the Kato-Ogawa theory.  
In Sec.7, we make a unified treatment of the BRS charges for both 
infinite and finite strings.  
The final section is devoted to discussion.
\vskip50pt
\Sec{Review of our previous paper}
We briefly review our previous work\cite{AN3} on the conformal-gauge 
two-dimensional quantum gravity coupled with $D$ scalar fields, which 
represent the coordinates of an infinite string.
\par
In the conformal gauge, the gravitational field $g^{\mu\nu}$ is 
parametrized as
\begin{eqnarray}
&& g^{\mu\nu}=e^{-\theta}(\eta^{\mu\nu}+h^{\mu\nu})
\end{eqnarray}
with $\eta_{\mu\nu}h^{\mu\nu}=0\ (\eta_{00}=-\eta_{11}=1,\;
\eta_{01}=0)$.
Then the conformal degree of freedom, $\theta$, disappears from the 
action. Corresponding to the fact that $h^{\mu\nu}$ is a traceless 
symmetric tensor, the B field $\tb_{\mu\nu}$ and the FP antighost 
$\barc_{\mu\nu}$ are also traceless symmetric tensors, while the FP 
ghost $c^\mu$ is a vector.
It is, therefore, convenient to rewrite a traceless symmetric tensor, 
which is generically denoted by $X_{\mu\nu}$, into a vectorlike quantity
\begin{eqnarray}
&&X^\lambda\equiv {1\over\sqrt2}\xi^{\lambda\mu\nu}X_{\mu\nu},
\end{eqnarray}
where $\xi^{\mu\nu\lambda}$ is a constant, totally symmetric rank-3 
tensorlike quantity, defined by $\xi^{\mu\nu\lambda}=1$ for 
$\mu+\nu+\lambda=$even, $\xi^{\mu\nu\lambda}=0$ otherwise.  
According to \eqno(2,2), 
we introduce $h_\lambda,\ \tb^\lambda,\ \barc^\lambda$.
\par
Let $\phi_M\ (M=0,\,1,\,\ldots\,,\,D-1; \ 
\eta^{MN}=(-1)^{\delta_{M0}}\delta_{MN})$ be scalar fields, which 
represent the coordinates of a string.   
The BRS transforms of the field operators are as follows:
\begin{eqnarray}
&&\gdel h_\lambda=\sqrt2\xi_{\lambda\mu\nu}\partial^\mu c^\nu
   +\xi_{\lambda\mu\nu}\xi^{\mu\sigma\tau}h_\sigma\partial_\tau c^\nu
   -\partial_\nu(h_\lambda c^\nu)
   -{1\over\sqrt2}h_\lambda\xi^{\nu\sigma\tau}h_\nu
    \partial_\sigma c_\tau,
   \\
&&\gdel c^\lambda=-c^\sigma\partial_\sigma c^\lambda, \\
&&\gdel \barc^\lambda = i\tb^\lambda,\\
&&\gdel \tb^\lambda=0,\\ 
&&\gdel \phi_M=-c^\sigma\partial_\sigma \phi_M.
\end{eqnarray}
The BRS-invariant Lagrangian density is given by 
\begin{eqnarray}
&&\hspace*{-20pt}
\lag=-{1\over\,2\,}\tb^\lambda h_\lambda
       -{i\over\,2\,}\barc^\lambda \gdel{ h_\lambda }
       +{1\over\,2\,}(1-\det h^{\sigma\tau})^{-1/2}
        (\eta^{\mu\nu}+h^{\mu\nu})
        \partial_\mu\phi_M\cdot\partial_\nu\phi^M.
\end{eqnarray}
\par
The field equations are as follows:
\begin{eqnarray}
&&h_\mu=0,\\
&&\tb^\mu= 
  -i[\xi_{\sigma\tau\rho}\xi^{\rho\mu\lambda}
     \barc^\sigma\partial_\lambda c^\tau
  + \partial_\sigma\barc^\mu\cdot c^\sigma]
  + {1\over\sqrt2}\xi^{\mu\sigma\tau}\partial_\sigma\phi_M
     \cdot\partial_\tau\phi^M, \\
&&\xi_{\lambda\mu\nu}\partial^\mu X^\nu=0 \quad \hbox{for\ } 
     X^\nu=c^\nu,\ \barc^\nu,\  \tb^\nu, \\ 
&&\square \phi_M=0,
\end{eqnarray}
where \eqno(2,11) for $X^\nu=\tb^\nu$ follows from \eqno(2,10).
\par
From the canonical (anti)commutation relations and the field equations 
given above, we can explicitly calculate the two-dimensional 
(anti)commutation relations.  We find
\begin{eqnarray}
&&\{ c^\rho(x),\;\barc^\lambda(y)\}
=\sqrt2\xi^{\rho\lambda\nu}\partial_\nu D(x-y), \\
&&[\phi_M(x),\;\phi^N(y)]=i\delta_M{}^N D(x-y),
\end{eqnarray}
where $D(x)\equiv -{1\over\,2\,}\ep(x^0)\theta(x^2)$.  
Hence if \eqno(2,10) were discarded by regarding it merely as the 
definition of $\tb^\mu$, then the model considered would be a free 
field theory.
This is not the right way, however, because the nonlinearity of 
\eqno(2,10) is the origin of anomaly.
\par
The two-dimensional commutation relations involving the B field can be 
calculated by using \eqno(2,10) together with \eqno(2,13) and 
\eqno(2,14).
The results are much simplified if we employ light-cone coordinates
$x^\pm\equiv(x^0\pm x^1)/\sqrt{2}$, with which $\xi_{\mu\nu\lambda}=0$ 
except $\xi_{+++}=\xi_{---}=\sqrt{2}$.
Then \eqno(2,10) and \eqno(2,11) reduce to 
\begin{eqnarray}
&&\tb^\pm=-i(2\barc^\pm\partial_\pm c^\pm
              +\partial_\pm\barc^\pm\cdot c^\pm)
           +\partial_\pm\phi_M\cdot\partial_\pm\phi^M 
            \equiv \tcalT^{\pm}, \\
&&\partial_\mp X^\pm=0 \quad 
  \hbox{for \ } X^\pm=c^\pm,\ \barc^\pm,\ \tb^\pm,
\end{eqnarray}
respectively.  Furthermore, since
\begin{eqnarray}
&&\partial_\pm D(x)=-{1\over\,2\,}\delta(x^\pm),
\end{eqnarray}
\eqno(2,13) and \eqno(2,14) reduce to
\begin{eqnarray}
&&\{c^\pm(x),\;\barc^\pm(y)\}=-\delta(x^\pm-y^\pm), \\
&&[\partial_\pm\phi_M(x),\;\phi^N(y)]
   =-{i\over\,2\,}\delta_M{}^N\delta(x^\pm-y^\pm),
\end{eqnarray}
respectively.  
Except for $[\phi_M(x),\;\phi^N(y)]$, the $+$ coordinate and 
the $-$ one never coexist in the right-hand side.
From \eqno(2,15) together with \eqno(2,18) and \eqno(2,19), 
we obtain
\begin{eqnarray}
[ \tb^\pm(x),\;c^\pm(y) ]
 &=&-i[c^\pm(x)\delta'(x^\pm-y^\pm)+2\partial_\pm c^\pm(x)
          \cdot\delta(x^\pm-y^\pm)], \\
{[} \tb^\pm(x),\;\barc^\pm(y) ]
 &=&i[\barc^\pm(x)+\barc^\pm(y)]\delta'(x^\pm-y^\pm)\nonumber\\
 &=&i[2\barc^\pm(x)\delta'(x^\pm-y^\pm)
      +\partial_\pm\barc^\pm(x)\cdot\delta(x^\pm-y^\pm)],\\
{[} \tb^\pm(x),\;\phi_M(y) ]
 &=&-i\partial_\pm\phi_M(x)\cdot\delta(x^\pm-y^\pm), \\
{[} \tb^\pm(x),\;\tb^\pm(y) ]
 &=&i[\tb^\pm(x)+\tb^\pm(y)]\delta'(x^\pm-y^\pm).
\end{eqnarray}
The totality of \eqno(2,18)$\sim$\eqno(2,23) constitutes the field 
algebra of the conformal-gauge two-dimensional quantum gravity.
\par
The representation of this algebra in terms of state vectors is given 
by constructing all (truncated\foot(a,{Truncation means to drop the 
contributions from vacuum intermediate states.})$\!\!$) $n$-point 
Wightman functions explicitly.
All 1-point Wightman functions vanish.  Nonvanishing 2-point Wightman 
functions are\foot(b,{Without differentiation in {\eqno(2,25)}, we must 
introduce an infrared cutoff.})
\begin{eqnarray}
&&\vwightman{\barc^\pm(x_1)c^\pm(x_2)}
  =\vwightman{c^\pm(x_1)\barc^\pm(x_2)}
  ={i\over2\pi}{1\over x_1{}^\pm-x_2{}^\pm-i0},\\
&&\partial_\pm{}^{x_1}\vwightman{\phi_M(x_1)\phi^N(x_2)}
 =-{1\over4\pi}\delta_M{}^N{1\over x_1{}^\pm-x_2{}^\pm-i0}.
\end{eqnarray}
Nonvanishing truncated $n$-point Wightman functions are those which 
consist of $(n-2)$ $\tb^\pm$'s and of either $c^\pm$ and $\barc^\pm$ 
or two $\phi_M$'s.  
For simplicity, we present the expressions for those of particular
orderings:\foot(c,{Those of the other orderings are obtained by 
changing $-i0$ into $+i0$ appropriately {(and the overall sign is 
changed if $c$ and $\barc$ are exchanged)}.})
\begin{eqnarray}
&&\hspace*{-25pt}
\vwightman{c^\pm(x_1)\tb^\pm(x_2)\cdots\tb^\pm(x_{n-1})\barc^\pm(x_n)} 
  \nonumber \\
&&=-i^{-n}\sum_{P(j_2,\,\cdots,\,j_{n-1})}^{(n-2)!}
  \left[\prod_{s=2}^{n-1}(\Lpartial{j_s}+2\Rpartial{j_s})\right]
  \wightman{1,\,j_2}^\pm \wightman{j_2,\,j_3}^\pm  \nonumber \\ 
&&\hspace*{180pt}
\cdots  \wightman{j_{n-2},\,j_{n-1}}^\pm\wightman{j_{n-1},\,n}^\pm, \\
&&\hspace*{-25pt}
\vwightman{\phi_M(x_1)\tb^\pm(x_2)\cdots\tb^\pm(x_{n-1})\phi^N(x_n)} 
  \nonumber\\
&&\quad
 =-{i^{-n+1} \over 2 }\delta_M{}^N
   \sum_{P(j_2,\,\cdots,\,j_{n-1})}^{(n-2)!}
   \left[\prod_{s=2}^{n-2}\Rpartial{j_s}\right]
   \wightman{1,\,j_2}^\pm\wightman{j_2,\,j_3}^\pm \nonumber \\
&&\hspace*{180pt} 
\cdots\wightman{j_{n-2},\,j_{n-1}}^\pm
   \wightman{j_{n-1},\,n}^\pm  \quad (n\geqq3), \hspace*{20pt}
\end{eqnarray}
where $P(j_2,\,\cdots,\,j_{n-1})$ is a permutation of 
$(2,\,3,\,\cdots,\, n-1)$, $\Lpartial{j}$ and $\Rpartial{j}$ denote 
differentiations with respect to $x_j{}^\pm$ acting only on the left 
factor involving $x_j{}^\pm$ and only on the right one, respectively, 
and
\begin{eqnarray}
&& \wightman{j,\,k}^\pm
   \equiv {i\over 2\pi}{1\over x_j{}^\pm-x_k{}^\pm-(k-j)i0}.
\end{eqnarray}
\par
A composite-field operator is a product of field operators of the same 
spacetime point.   The Wightman function involving a composite field is 
obtained from the (nontruncated) Wightman function by setting the 
spacetime coordinates of consecutive field operators coincident and 
by discarding the infinities which appear as a consequence in such 
a way that the result be independent of the ordering of the constituent 
fields of the composite field. 
The latter procedure is called ``generalized normal product'' 
because it reduces to Wick's normal product in the free-field case.
\par
The representation of the field algebra in terms of Wightman functions 
is, of course, consistent with all two-dimensional (anti)commutation 
relations and also with all linear field equations (including 
\eqno(2,16) for $X^\pm=\tb^\pm$).
However, it is {\it not\/} consistent with the B-field equation 
\eqno(2,15).  Indeed, we have
\begin{eqnarray}
&&\hspace*{-20pt}\vwightman{\tb^\pm(x_1)\tb^\pm(x_2)}=0, \\
&&\hspace*{-20pt}\vwightman{\tb^\pm(x_1)\tcalT^\pm(x_2)}
=\vwightman{\tcalT^\pm(x_1)\tcalT^\pm(x_2)}
=-{1\over\,2\,}(D-26)[\partial_\pm{}^{x_1}\wightman{1,2}^\pm]^2,
\hspace*{30pt}
\end{eqnarray}
where $\tcalT^\pm$ denotes the right-hand side of \eqno(2,15).
Thus the B-field equation \eqno(2,15), but not \eqno(2,16) for 
$X^\pm=\tb^\pm$, is {\it violated at the representation level.\/}  
We call this situation ``field-equation anomaly''.
The field-equation anomaly disappears for $D=26$ in the conformal 
gauge, but this property does not remain valid in the de Donder 
gauge.\cite{AN1}
\par
The BRS Noether current is given by
\begin{eqnarray}
&&j_b{}^\mp=-i\barc^\pm c^\pm \partial_\pm c^\pm
            -c^\pm \partial_\pm \phi_M\cdot \partial_\pm \phi^M,
\end{eqnarray}
and the corresponding BRS charge is defined by
\begin{eqnarray}
&&Q_b={1\over\sqrt{2}}\int dx^1[j_b{}^-(x^+)+j_b{}^+(x^-)].
\end{eqnarray}
{\it This\/} BRS charge is, however, anomalous.  
We rewrite \eqno(2,31) as
\begin{eqnarray}
&& j_b^{\mp}=\hj{}_b{}^\mp + (\tb^\pm - \tcalT^\pm)c^\pm \\
\noalign{\noindent with}
&& \hj{}_b{}^\mp
 \equiv -\tb^\pm c^\pm +i\barc^\pm c^\pm \partial_\pm c^\pm.
\end{eqnarray}
Because of \eqno(2,15), $\hj{}_b{}^\mp$ is the same operator as 
$j_b{}^\mp$ at the operator level.  
They no longer coincide, however, at the representation level because 
of field-equation anomaly. 
The BRS charge defined by $\hj{}_b{}^\mp$, i.e.,
\begin{eqnarray}
&&\skew3\hat Q{}_b={1\over\sqrt{2}}\int dx^1
  [\hj{}_b{}^-(x^+) + \hj{}_b{}^+(x^-)],
\end{eqnarray}
is {\it free of anomaly for any value of $D$\/}.
Of course, any vacuum expectation value involving 
$\skew3\hat Q{}_b{}^2$ is zero independently of $D$ (see Sec.7).
\vskip50pt
\Sec{Closed String}
In this section, we consider how the formulae presented in Sec.2 are 
modified if the string is not of infinite length but a finite ring.
\par
Let the length of the string be $2\pi$.  That is, every function of 
$x^\mu$ must be periodic in $x^1$ with a period $2\pi$.  It is well 
known in the curved-spacetime quantum field theory\cite{Birrell-Davies} 
how to treat field operators and Green's functions in such a situation.
\par
The two-dimensional commutator D-function is modified into
\begin{eqnarray}
&&D_f(x)\equiv -{1\over\,2\,}\ep(x^0)\sum_{m=-\infty}^{\infty}
                         \theta( (x^0)^2-(x^1-2\pi m)^2 ), \\
\noalign{\noindent whence}
&&\partial_\pm D_f(x)=-{1\over\,2\,}\sum_{m=-\infty}^{\infty}
                                      \delta(x^\pm-\sqrt{2}\pi m).
\end{eqnarray}
Accordingly, \eqno(2,18)$\sim$\eqno(2,23) are replaced by
\begin{eqnarray}
&&\{ c^\pm(x), \; \barc^\pm(y) \}
  =2\partial_\pm D_f(x-y), \\
&&[\partial_\pm\phi_M(x),\; \phi^N(y)]
  =i\delta_M{}^N\partial_\pm D_f(x-y), \\
&&[\tb^\pm(x),\; c^\pm(y)]
  =2i[c^\pm(x)\partial_\pm + 2\partial_\pm c^\pm(x)]
      \partial_\pm D_f(x-y), \\
&&[\tb^\pm(x),\; \barc^\pm(y)]
  =-2i[2\barc^\pm(x)\partial_\pm 
      +\partial_\pm\barc^\pm(x)]\partial_\pm D_f(x-y), \\
&&[\tb^\pm(x),\;\phi_M(y)]
  =2i\partial_\pm\phi_M(x)\cdot\partial_\pm D_f(x-y),\\
&&[\tb^\pm(x),\;\tb^\pm(y)]
  =-2i[\tb^\pm(x)+\tb^\pm(y)](\partial_\pm)^2 D_f(x-y).
\end{eqnarray}
\par
The Wightman functions are, therefore, obtained from 
\eqno(2,24)$\sim$\eqno(2,27) by the replacement
\begin{eqnarray}
\hspace*{-30pt}&& 
\partial_\pm \Dp(x)=-{1\over 4\pi}{1\over x^\pm-i0}  
  \ \ \Longrightarrow \ \ 
  \partial_\pm D_f^{(+)}(x)
  \equiv -{1\over 4\pi}\sum_{m=-\infty}^{\infty}
        {1\over x^\pm-\sqrt{2}\pi m -i0}.
\end{eqnarray}
The proof of the BRS invariance for Wightman functions, done in our 
previous paper\cite{AN3}, can be straightforwardly extended to the 
present case.
\par
The summation in \eqno(3,9) can be explicitly carried out to obtain
\begin{eqnarray}
\partial_\pm D_f^{(+)}(x)
 &=& -{1\over 4\pi} {1\over\sqrt2}\cot\Bigg({x^\pm\over\sqrt2}-i0\Bigg) 
     \nonumber \\
 &=& -{i\over 2\sqrt{2}\pi}\Bigg( {1\over\,2\,} 
     + \sum_{n=1}^\infty e^{-in\sqrt{2}x^\pm}\Bigg).
\end{eqnarray}
Hence, we have
\begin{eqnarray}
&&\partial_\pm D_f(x)
  =i^{-1}[\partial_\pm D_f^{(+)}(x)+\partial_\pm D_f^{(+)}(-x)]
  =-{1\over 2\sqrt{2}\pi}\sum_{n=-\infty}^\infty e^{-in\sqrt{2}x^\pm}.
\end{eqnarray}
\par
It is interesting to introduce mode expansions.  
For $X^\lambda=c^\lambda,\ \barc^\lambda,\ \tb^\lambda$, we write
\begin{eqnarray}
&& X^\pm(x)={1\over\sqrt{2\pi}}\sum_{n=-\infty}^\infty 
\bX_n^\pm e^{-in\sqrt{2}x^\pm}
\end{eqnarray}
with $\bX_n^\pm= \bc_n^\pm,\ \barbc_n^\pm,\ \tbb_n^\pm$.   
Then \eqno(3,3), \eqno(3,5),
\eqno(3,6) and \eqno(3,8) are rewritten in terms of mode operators:
\begin{eqnarray}
&& \{ \bc_n^\pm,\; \barbc_m^\pm \}
    =-\sqrt{2}\,\delta_{n,\,-m}, \\
&& [\tbb_n^\pm,\; \bc_m^\pm]
    =-\sqrt{2\over\,\pi\,}(2n+m)\bc_{n+m}^\pm, \\
&& [\tbb_n^\pm,\; \barbc_m^\pm]
    =\sqrt{2\over\,\pi\,}(n-m)\barbc_{n+m}^\pm, \\
&& [\tbb_n^\pm,\; \tbb_m^\pm]
    =\sqrt{2\over\,\pi\,}(n-m)\tbb_{n+m}^\pm,
\end{eqnarray}
respectively.  Of course, $\bX_n^\pm$ and $\bY_m^\mp$ (anti)commute.
\vfill\eject
\Sec{Kato-Ogawa string theory}
Kato and Ogawa\cite{Kato-Ogawa} presented the BRS formalism of the open 
string of length $\pi$.
Since their starting Lagrangian density is not the same as ours, we 
here start with their formulae of the mode expansions of field 
operators.  
To make the comparison easier, we translate their notation into ours 
in the following way.\vfoot(d,{See first two formulae of 
Kato-Ogawa's {\eqno(2,13)}.})
\begin{eqnarray*}
&&
\begin{array}{ll}
\hbox{2-dimensional coordinates:\ } 
& \sigma \ \Rightarrow \ x^1, \quad \tau \ \Rightarrow \ x^0\ ; \\
\noalign{\vskip5pt}
\hbox{2-dimensional indices:\ }
& a,\ b,\ \cdots \ \Rightarrow \ \mu,\ \nu,\ \cdots\ ; \\
\noalign{\vskip5pt}
\hbox{string component indices:\ }
& \mu,\ \nu\ \Rightarrow \ M,\ N \ ; \\
\noalign{\vskip5pt}
\hbox{string-space metric:\ }
&g^{\mu\nu} \ \Rightarrow \ -\eta^{MN}\ ; \\
\noalign{\vskip5pt}
\hbox{field operators:\ }
&{1\over\sqrt\kappa}X_\mu \ \Rightarrow \ \phi_M,\quad 
 c^a\ \Rightarrow \ c^\mu, \\
\noalign{\vskip5pt}
&\barc_0\ \Rightarrow \ -{1\over\sqrt2}\barc^1,\quad
  \barc_1\ \Rightarrow \ -{1\over\sqrt2}\barc^0, \\
\noalign{\vskip5pt}
&B_0\ \Rightarrow\ -{1\over\sqrt2}\tb^1,\quad
  B_1\ \Rightarrow\ -{1\over\sqrt2}\tb^0,\  \hbox{where $B_a$ is }\\
\noalign{\vskip5pt}
& \hbox{the $B_a$ before field redefinition$^d$ is made;}\\
\noalign{\vskip5pt}
\hbox{zero-mode operators:\ }
&{\sqrt{\pi\over\kappa}}q_0^\mu\ \Rightarrow\  \bq_0^M,\quad
 {\sqrt{\kappa\over\pi}}p_0^\mu\ \Rightarrow\ \bp_0^M, \quad  \\
\noalign{\vskip5pt}
& c_0\ \Rightarrow \ \bc_0,\quad 
  \barc_0 \ \Rightarrow\ -{1\over\sqrt2}\barbc_0; \\ 
\noalign{\vskip5pt}
\hbox{nonzero-mode operators:\ }
& a_n{}^\mu \ \Rightarrow\ \ba_n{}^M, \\ 
& c_n\ \Rightarrow\ \bc_n,\quad 
  \barc_n\ \Rightarrow\ -{1\over\sqrt2}\barbc_n. 
\end{array}
\end{eqnarray*}
In our notation, their mode expansion formulae [Kato-Ogawa's 
\eqno(2,18)] are translated into
\begin{eqnarray}
&&\phi^M(x)
  ={1\over\sqrt{\pi}}\bq_0{}^M+{1\over\sqrt{\pi}}\bp_0{}^Mx^0 
   \nonumber \\
&&\hspace*{50pt}
   +{1\over\sqrt{\pi}}\sum_{n=1}^\infty{1\over\sqrt{n}}
    (\ba_n{}^M e^{-inx^0}+\ba_n{}^M{}^\dagger e^{inx^0})\cos{nx^1}, \\
&&c^0(x)
  ={1\over\sqrt{\pi}}\bc_0+{1\over\sqrt{\pi}}\sum_{n=1}^\infty
    (\bc_n e^{-inx^0}+\bc_n{}^\dagger e^{inx^0})\cos{nx^1},\\
&&c^1(x)
  =-{i\over\sqrt{\pi}}\sum_{n=1}^\infty
    (\bc_n e^{-inx^0}-\bc_n{}^\dagger e^{inx^0})\sin{nx^1},\\
&&\barc^1(x)
  =-{i\over\sqrt{\pi}}\sum_{n=1}^\infty
    (\barbc_n e^{-inx^0}-\barbc_n{}^\dagger e^{inx^0})\sin{nx^1},\\
&&\barc^0(x)
  ={1\over\sqrt{\pi}}\barbc_0+{1\over\sqrt{\pi}}\sum_{n=1}^\infty
    (\barbc_n e^{-inx^0}+\barbc_n{}^\dagger e^{inx^0})\cos{nx^1}
\end{eqnarray}
with
\begin{eqnarray}
&&[\bp_0{}^M,\;\bq_0{}^N]=-i\eta^{MN},\qquad
  [\ba_n{}^M,\;\ba_m{}^N{}^\dagger]=\eta^{MN}\delta_{nm},\\
&&\{\bc_0,\;\barbc_0\}=-\sqrt{2}, \qquad
  \{\bc_n,\;\barbc_m{}^\dagger\}
  =\{\bc_n{}^\dagger,\;\barbc_m\}=-\sqrt{2}\delta_{nm},
\end{eqnarray}
others being zero.  
We rewrite \eqno(4,2)$\sim$\eqno(4,5) as
\begin{eqnarray}
c^\pm(x)
&=&{1\over\sqrt{2\pi}}\Bigg[ \bc_0 
  +\sum_{n=1}^\infty(\bc_ne^{-in\sqrt{2}x^\pm}
  +\bc_n{}^\dagger e^{in\sqrt{2}x^\pm})\Bigg] \nonumber\\
&=&{1\over\sqrt{2\pi}}\sum_{n=-\infty}^\infty 
   \bc_n e^{-in\sqrt{2}x^\pm},\\
\barc^\pm(x)
&=&{1\over\sqrt{2\pi}}\Bigg[ \barbc_0
  +\sum_{n=1}^\infty(\barbc_n e^{-in\sqrt{2}x^\pm}
  +\barbc_n{}^\dagger e^{in\sqrt{2}x^\pm})\Bigg] \nonumber\\
&=&{1\over\sqrt{2\pi}}\sum_{n=-\infty}^\infty 
   \barbc_n e^{-in\sqrt{2}x^\pm},
\end{eqnarray}
where $\bc_{-n}\equiv\bc_n{}^\dagger,\ 
\barbc_{-n}\equiv\barbc_n{}^\dagger$, and 
$\{\bc_n,\;\barbc_m\}=-\sqrt{2}\delta_{n,\,-m}$.
Compared with \eqno(3,12), we note that the mode operators for the open 
string has no $\pm$ index.
From \eqno(4,8) and \eqno(4,9), we obtain
\begin{eqnarray}
&&\{c^\pm(x),\;\barc^\pm(y)\}=-{1\over\sqrt{2}\pi}
 \Bigg[1+2\sum_{n=1}^\infty \cos{n\sqrt{2}(x^\pm-y^\pm)} \Bigg], \\
&&\{c^\pm(x),\;\barc^{\mp}(y)\}=-{1\over\sqrt{2}\pi}
 \Bigg[1+2\sum_{n=1}^\infty \cos{n\sqrt{2}(x^\pm-y^\mp)} \Bigg].
\end{eqnarray}
The nonvanishing of \eqno(4,11) is due to translational noninvariance.
\par
In the Fourier expansion formula
\begin{eqnarray}
&&{1-\lambda^2\over 1-2\lambda \cos\alpha +\lambda^2}
= 1+2\sum_{n=1}^\infty \lambda^n\cos{n\alpha} \qquad (|\lambda|<1),
\end{eqnarray}
we take the limit $\lambda\ \to\ 1$ ; we then see that the left-hand 
side of \eqno(4,12) tends to
\begin{eqnarray}
&& 2\pi \sum_{m=-\infty}^\infty \delta(\alpha-2\pi m). 
\end{eqnarray}
Hence \eqno(4,10) and \eqno(4,11) are rewritten as
\begin{eqnarray}
&& \{c^\pm(x),\;\barc^\pm(y)\}
   =-\sum_{m=-\infty}^\infty \delta(x^\pm-y^\pm-\sqrt{2}\pi m), \\
&& \{c^\pm(x),\;\barc^\mp(y)\}
   =-\sum_{m=-\infty}^\infty \delta(x^\pm-y^\mp-\sqrt{2}\pi m),  
\end{eqnarray}
respectively.
\par
Next, from \eqno(4,1) with \eqno(4,6), we obtain
\begin{eqnarray}
\hspace*{-30pt}
&&[\phi_M(x),\;\phi^N(y)]=-{i\over\pi}\delta_M{}^N
\Bigg\{ x^0-y^0 
  +{1\over\,2\,}\sum_{n=1}^\infty{1\over n}\bigg(
   \sin{n\sqrt{2}(x^+-y^+)}
\nonumber \\
\hspace*{-30pt}
&&\hspace*{50pt}
+\sin{n\sqrt{2}(x^--y^-)} 
  +\sin{n\sqrt{2}(x^+-y^-)}+\sin{n\sqrt{2}(x^--y^+)} \bigg) \Bigg\}.
\end{eqnarray}
Using the formula
\begin{eqnarray}
&&\sum_{n=1}^\infty{\sin{n\alpha}\over n}
  ={1\over\,2\,}(\pi-\alpha)+\pi\bigg[\, {\alpha\over 2\pi} \,\bigg],
\end{eqnarray}
where $[\,r\,]$ denotes the largest integer not greater than $r$, 
we obtain
\begin{eqnarray}
&&[\phi_M(x),\;\phi^N(y)]
=-{1\over\,2\,}i\delta_M{}^N\Bigg\{ 2 
+ \bigg[\, {x^+ - y^+ \over \sqrt{2}\pi} \,\bigg] \nonumber \\
&&\hspace*{100pt}
+ \bigg[\, {x^- - y^- \over \sqrt{2}\pi} \,\bigg] 
+ \bigg[\, {x^+ - y^- \over \sqrt{2}\pi} \,\bigg]
+ \bigg[\, {x^- - y^+ \over \sqrt{2}\pi} \,\bigg] \Bigg\}, \\
\noalign{\noindent and, therefore,}
&&[\partial_\pm\phi_M(x),\;\phi^N(y)]
=-{1\over\,2\,}i\delta_M{}^N \sum_{m=-\infty}^\infty
\big[\delta(x^\pm-y^\pm-\sqrt{2}\pi m) \nonumber \\
&&\hspace*{190pt} + \ \delta(x^\pm-y^\mp-\sqrt{2}\pi m)\big].
\end{eqnarray}
\par
Now, we consider the B field.  Kato and Ogawa first made field 
redefinition [Kato-Ogawa's \eqno(2,13)] and then wrote down the field 
equations in terms of the redefined fields [Kato-Ogawa's \eqno(2,16)].
We should, therefore, restore the field equation for the original 
B field from those formulae.  We then find
\begin{eqnarray}
&&\tb^\pm=\partial_\pm\phi_M\cdot\partial_\pm\phi^M
 -i(2\barc^\pm\partial_\pm c^\pm + \partial_\pm\barc^\pm\cdot c^\pm)
\end{eqnarray}
in our notation.  
As expected, \eqno(4,20) is identical with \eqno(2,15).
\par
We calculate the commutators involving the B field by using \eqno(4,10),
\eqno(4,11), \eqno(4,14), \eqno(4,15) and \eqno(4,19).  We find
\begin{eqnarray}
&&[\tb^\pm(x),\; c^\pm(y)]
  =-i[c^\pm(x)\partial_\pm+2\partial_\pm c^\pm(x)]
    \sum_{m=-\infty}^\infty\delta(x^\pm-y^\pm-\sqrt{2}\pi m), \\
&&[\tb^\pm(x),\; c^\mp(y)]
  =-i[c^\pm(x)\partial_\pm+2\partial_\pm c^\pm(x)]
    \sum_{m=-\infty}^\infty\delta(x^\pm-y^\mp-\sqrt{2}\pi m); \\
&&[\tb^\pm(x),\; \barc^\pm(y)]
  =i[2\barc^\pm(x)\partial_\pm+\partial_\pm \barc^\pm(x)]
    \sum_{m=-\infty}^\infty\delta(x^\pm-y^\pm-\sqrt{2}\pi m), \\
&&[\tb^\pm(x),\; \barc^\mp(y)]
  =i[2\barc^\pm(x)\partial_\pm+\partial_\pm \barc^\pm(x)]
    \sum_{m=-\infty}^\infty\delta(x^\pm-y^\mp-\sqrt{2}\pi m); \\
&&[\tb^\pm(x),\; \phi_M(y)]
  =-i\partial_\pm\phi_M(x)\sum_{m=-\infty}^\infty
    [\delta(x^\pm-y^\pm-\sqrt{2}\pi m) \nonumber \\
&&\hspace*{200pt}
  +\ \delta(x^\pm-y^\mp-\sqrt{2}\pi m)];\\
&&[\tb^\pm(x),\;\tb^\pm(y)]
  =i(\tb^\pm(x)+\tb^\pm(y))\sum_{m=-\infty}^\infty
    \delta'(x^\pm-y^\pm-\sqrt{2}\pi m), \\
&&[\tb^\pm(x),\;\tb^\mp(y)]
  =i(\tb^\pm(x)+\tb^\mp(y))\sum_{m=-\infty}^\infty
    \delta'(x^\pm-y^\mp-\sqrt{2}\pi m).
\end{eqnarray}
In terms of mode operators, we have
\begin{eqnarray}
&&[\tbb_n,\;\bc_m]=-\sqrt{2\over\,\pi\,}\, (2n+m)\bc_{n+m}, \\
&&[\tbb_n,\;\barbc_m]=\sqrt{2\over\,\pi\,}\, (n-m)\barbc_{n+m}, \\
&&[\tbb_n,\;\tbb_m]=\sqrt{2\over\,\pi\,}\, (n-m)\tbb_{n+m},
\end{eqnarray}
where $\tbb_n$ is the mode operator of the B field defined through
\begin{eqnarray}
&& \tb^\pm(x)={1\over\sqrt{2\pi}}\sum_{n=-\infty}^\infty
   \tbb_n e^{-in\sqrt{2}x^\pm}.
\end{eqnarray}
\par
One should note the parallelism between the above formulae and those 
for the closed string.
\vskip50pt
\Sec{FP-ghost vacuum structure of the Kato-Ogawa theory}
Kato and Ogawa\cite{Kato-Ogawa} introduced the vacuum structure 
in a rather artificial way, especially for the FP-ghost.  
They introduced two FP-ghost vacua $|\,+\,\rangle$ and $|\,-\,\rangle$ 
such that $c_0|\,+\,\rangle=0$,\ $\barc_0|\,+\,\rangle=|\,-\,\rangle$,\ 
$\barc_0|\,-\,\rangle=0$,\ $c_0|\,-\,\rangle=|\,+\,\rangle$, 
where $\{c_0,\;\barc_0\}=1$.
Since $c_0{}^\dagger=c_0$ and $\barc_0{}^\dagger=\barc_0$, both vacua 
are of zero norm.  To overcome this trouble, they assumed 
$\langle\,+\,|\,-\,\rangle=\langle\,-\,|\,+\,\rangle=1$ and introduced 
an indefinite metric $\eta=c_0+\barc_0$ by hand so that 
$\eta |\,+\,\rangle=|\,-\,\rangle$ and 
$\eta |\,-\,\rangle=|\,+\,\rangle$.
This procedure is {\it not admissible\/}, however, because the 
introduction of $\eta$ violates the operator hermitian conjugation 
at the representation level.  Indeed, in the presence of $\eta$, 
the original action is no longer hermitian in the sense of 
Kato-Ogawa's inner product.
\par
More naturally, we should start with the unique vacuum $\rvac$ with 
positive norm,
\begin{eqnarray}
&&\langle\,0\,|\,0\,\rangle=1.
\end{eqnarray}
Then the trouble encountered is how to calculate 
$\vwightman{\bc_0\barbc_0}$.
This problem can be resolved in the following way.
\par
From the consideration made in the present paper, it is now 
straightforward to calculate all Wightman functions in the Kato-Ogawa 
theory.  For example, we have
\begin{eqnarray}
&&\vwightman{c^\pm(x_1)\barc^\pm(x_2)}
  ={i\over2\pi}\sum_{m=-\infty}^\infty
   {1\over x_1{}^\pm-x_2{}^\pm -\sqrt{2}\pi m -i0}, \\
&&\vwightman{c^\pm(x_1)\barc^\mp(x_2)}
  ={i\over2\pi}\sum_{m=-\infty}^\infty
   {1\over x_1{}^\pm-x_2{}^\mp -\sqrt{2}\pi m -i0},
\end{eqnarray}
We note from \eqno(4,8) and \eqno(4,9) that
\begin{eqnarray}
&&\int_0^\pi dx^1 \,(c^+(x)+c^-(x))=\sqrt{2\pi}\,\bc_0, \\
&&\int_0^\pi dx^1 \,(\barc^+(x)+\barc^-(x))=\sqrt{2\pi}\,\barbc_0.
\end{eqnarray}
From \eqno(5,2)$\sim$\eqno(5,5), we obtain
\begin{eqnarray}
&& \hspace*{-55pt}
\vwightman{\bc_0 \barbc_0}\nonumber \\
&& \hspace*{-50pt}
={i\over 4\pi^2}\int_0^\pi dx^1\int_0^\pi dy^1 \sum_{m=-\infty}^\infty
\Bigg[ {1\over x^+-y^+-\sqrt{2}\pi m -i0}
      +{1\over x^--y^--\sqrt{2}\pi m-i0}  \nonumber \\
&&\hspace*{100pt}
      +{1\over x^+-y^--\sqrt{2}\pi m -i0}
      +{1\over x^--y^+-\sqrt{2}\pi m-i0}\Bigg].
\end{eqnarray}
Because of the periodicity, we may set $x^0-y^0=0$. Then the real part 
of the integrand is seen to vanish.  Thus \eqno(5,6) reduces to
\begin{eqnarray}
\vwightman{\bc_0\barbc_0}
&=&-{1\over\sqrt{2}\pi}\int_0^\pi dx^1\int_0^\pi dy^1 \delta(x^1-y^1) 
   \nonumber\\
&=&-{1\over\sqrt{2}}.
\end{eqnarray}
Likewise, we have
\begin{eqnarray}
\vwightman{\barbc_0\bc_0}&=&-{1\over\sqrt{2}}.
\end{eqnarray}
Of course, the sum of \eqno(5,7) and \eqno(5,8) is consistent with
\eqno(4,7) and \eqno(5,1).
\par
From the above consideration, it is natural to conclude that
$\rvac$, $\bc_0\rvac$, $\barbc_0\rvac$, and 
$(\bc_0\barbc_0-\barbc_0\bc_0)\rvac$ are {\it four linearly independent 
states\/}.  The Kato-Ogawa vacua $|\,+\,\rangle$ and $|\,-\,\rangle$ 
are interpreted as \begin{eqnarray}
&&|\,+\,\rangle = \bc_0(\alpha\barbc_0+\beta)\rvac,\\
&&|\,-\,\rangle = \barbc_0(\alpha'\bc_0+\beta')\rvac,
\end{eqnarray}
where $\alpha$, $\beta$, $\alpha'$ $\beta'$ are arbitrary $c$-numbers.
Each of them is {\it not a single state\/}.
Indeed, the relation $\eta|\,\pm\,\rangle=|\,\mp\,\rangle$ holds only 
in the sense of the {\it subspace\/}.
\par
The FP-ghost number operator is given by
\begin{eqnarray}
iQ_c&=& {1\over\sqrt{2}}\int_0^\pi \,dx^1\,
[ \barc^+(x)c^+(x)+\barc^-(x)c^-(x) ] \nonumber\\
&=&{1\over\sqrt{2}}\sum_{n=-\infty}^\infty \barbc_{-n}\bc_n.
\end{eqnarray}
Adjusting the zero-point value, we redefine $Q_c$ by
\begin{eqnarray}
iQ_c&=&{1\over\sqrt{2}}\Bigg[{1\over\,2\,}(\barbc_0\bc_0-\bc_0\barbc_0)
 + \sum_{n=1}^\infty(\barbc_n{}^\dagger \bc_n 
 - \bc_n{}^\dagger\barbc_n)\Bigg].
\end{eqnarray}
The Kato-Ogawa vacua satisfy
\begin{eqnarray}
&& iQ_c|\,\pm\,\rangle=\pm{1\over\,2\,}|\,\pm\,\rangle.
\end{eqnarray}
From this result, Kato and Ogawa concluded that the FP-ghost numbers 
would be half-integers.
\par
From our standpoint, the genuine vacuum $\rvac$ is {\it not\/} an 
eigenstate of $iQ_c$.\foot(e,{Freeman and Olive\cite{Freeman-Olive} 
also adopted a vacuum which is not an eigenstate of $iQ_c$, but theirs 
is a linear combination of $|\,+\,\rangle$ and $|\,-\,\rangle$.  
In their formalism, therefore, no physical states have a definite 
FP-ghost number.}) We can, however, bypass the trouble caused by this 
fact in the following way.
\par
Let $P\,(=P^\dagger)$ be the projection operator to the subspace defined
by the totality of the states which can be constructed from $\rvac$ by 
using the mode operators other than $\bc_0$ and $\barbc_0$.
Then we introduce $PQ_cP$ instead of $Q_c$.  We find that $iPQ_cP$ is 
unbroken and has integral eigenvalues.
There is no anomalous feature for it.  We should also introduce $PQ_bP$ 
for the BRS charge in order to keep the relation between the BRS charge
and the FP-ghost number.
\vskip50pt
\Sec{Nilpotent BRS charge in the Kato-Ogawa theory}
In this section, we establish our main claim that it is possible to 
construct a BRS charge nilpotent for any value of $D$ in the Kato-Ogawa
string theory.
\par
First, we briefly review how Kato and Ogawa\cite{Kato-Ogawa} obtained 
their crucial result\foot(f,{Unfortunately, in their paper, 
the denominator factor is incorrectly written as $12$ instead of 
$24$.\vskip8pt})
\begin{eqnarray}
&&Q_B{}^2={2\over\,\pi\,}\Bigg[ 
{D-26\over 24}\sum_{n=1}^\infty n^3 \bc_n{}^\dagger \bc_n
-\bigg( {D-26\over24}-\alpha_0+1 \bigg) \sum_{n=1}^\infty n
 \bc_n{}^\dagger\bc_n \Bigg].
\end{eqnarray}
They define their BRS charge $Q_B$ in terms of the BRS {\it Noether\/} 
current. That is, with
\begin{eqnarray}
&&Q_b\equiv{1\over\sqrt{2}}\int_0^\pi\,dx^1\,
[j_b{}^-(x)+j_b{}^+(x)], \\
\noalign{\noindent where}
&&j_b{}^\mp=-i\barc^\pm c^\pm\partial_\pm c^\pm 
            -c^\pm\partial_\pm\phi_M\cdot\partial_\pm\phi^M,
\end{eqnarray}
which is the same as \eqno(2,31), 
$Q_B$ is defined by the normal-product 
form of $Q_b$ (more precisely, see below).  They express $Q_B$ as
\begin{eqnarray}
&&Q_B=L\bc_0+M\barbc_0+\tilde Q_B,
\end{eqnarray}
where $L$, $M$ and $\skew3\tilde Q_B$ involve neither $\bc_0$ nor 
$\barbc_0$.\foot(g,{Slight changes of notation should be understood.})
Since the normal-product forms of $M$ and $\skew3\tilde Q_B$ are the 
same as themselves, the difference between $Q_B$ and $Q_b$ arises only 
from $L$. That is, we can formally write
\begin{eqnarray}
&&Q_B=Q_b+K\bc_0,\\
\noalign{\noindent with}
&&K\equiv {D-2\over2\sqrt{\pi}}\sum_{n=1}^\infty n 
  + {1\over\sqrt{\pi}}\alpha_0,
\end{eqnarray}
where $\alpha_0$ is a regularization parameter of the Hamiltonian 
(interpreted as the zero intercept of the leading Regge trajectory).  
From our standpoint of generalized normal product stated in Sec.2, 
however, we should set $\alpha_0=0$.  Note that $PQ_BP=PQ_bP$.
\par
At the operator level, we, of course, have
\begin{eqnarray}
&&Q_b{}^2=0,
\end{eqnarray}
as can be straightforwardly verified by explicit calculation.  Hence naive
operator calculation yields $Q_B{}^2=-\sqrt{2}KM$.
\par
Thus it is impossible to derive \eqno(6,1) unambiguously by operator
calculation.  We emphasize that the reasonable derivation of \eqno(6,1) 
can be done only on the basis of the Fock representation of 
nonzero-mode operators.
That is, {\it \eqno(6,1) is a formula which holds not at the operator 
level but at the representation level.}
Indeed, the expression \eqno(6,1) can be obtained by calculating matrix
elements of $Q_B{}^2$ with respect to Fock states.  Especially, it is 
easy to see
\begin{eqnarray}
&&\hspace*{-50pt}
{1\over\,2\,}\vwightman{\barbc_n Q_B{}^2 \barbc_m{}^\dagger}
={2\over\,\pi\,}\bigg[ {D-26\over24}(n^3-n)+(\alpha_0-1)n\bigg]
 \delta_{n\,m} \qquad (n,\ m>0).
\end{eqnarray}
\par
Since \eqno(6,1) is a result not at the operator level but at the 
representation level, it can be changed by using the field-equation 
anomaly, as discussed at the end of Sec.2.  
Owing to \eqno(4,20), \eqno(6,2) {\it equals\/}
\begin{eqnarray}
&&\skew3\hat Q_b
={1\over\sqrt2}\int_0^\pi\,dx^1\,[\hj_b{}^-(x) + \hj_b{}^+(x)]
 \\
\noalign{\noindent with}
&&\hj_b{}^\mp=-\tb^\pm c^\pm+i\barc^\pm c^\pm\partial_\pm c^\pm
\end{eqnarray}
at the operator level.  We now demonstrate that
\begin{eqnarray}
&&\vwightman{\barbc_n (:\!\skew3\hat Q_b\!:)^2 \barbc_m{}^\dagger}=0 
  \qquad (n,\ m>0)
\end{eqnarray}
{\it independently of the value of $D$\/}.
\par
In terms of mode operators, $:\!\skew3\hat Q_b\!:$ is given by
\begin{eqnarray}
:\!\skew3\hat Q_b\!:
&=&-{1\over\sqrt{2}}\sum_{n=-\infty}^\infty :\tbb_{-n}\bc_n:
 +{1\over\sqrt{2\pi}}\sum_{n=-\infty}^\infty \sum_{m=-\infty}^\infty
 m\,:\barbc_{-n-m}\bc_n\bc_m: \nonumber \\
&=&-{1\over\sqrt2}\Bigg(\tbb_0\bc_0+\sum_{n=1}^\infty\tbb_n{}^\dagger 
   \bc_n
  +\sum_{n=1}^\infty \bc_n{}^\dagger \tbb_n\Bigg) \nonumber\\
&&+{1\over\sqrt{2\pi}}\Bigg[
-\bc_0\sum_{n=1}^\infty n(\barbc_n{}^\dagger\bc_n
                           +\bc_n{}^\dagger\barbc_n)
 +2\barbc_0\sum_{n=1}^\infty n\bc_n{}^\dagger \bc_n \nonumber \\
&&\hspace*{50pt}
+\sum_{n=1}^\infty \sum_{m=1}^\infty m(\barbc_{n+m}{}^\dagger \bc_n
 \bc_m
 -\bc_n{}^\dagger\bc_m{}^\dagger\barbc_{n+m}) \nonumber\\
&&\hspace*{50pt}
+\sum_{n=1}^\infty \sum_{m=1}^\infty (n+2m)(
  \barbc_n{}^\dagger\bc_m{}^\dagger\bc_{n+m}
 +\bc_{n+m}{}^\dagger\bc_m\barbc_n)\Bigg].
\end{eqnarray}
Accordingly, by using \eqno(4,29) together with
\begin{eqnarray}
&&\hspace*{-20pt}
\tbb_n\rvac=\{:\!\skew3\hat Q_b\!:\,,\;\barbc_n\}\rvac
 =:\!\skew3\hat Q_b\!:\barbc_n\rvac+\barbc_n:\!\skew3\hat Q_b\!:\rvac=0 
\qquad (n>0),
\end{eqnarray}
where $:\!\skew3\hat Q_b\!:\rvac=0$ holds as is shown at the end of 
Sec.7, we have
\begin{eqnarray}
&&:\!\skew3\hat Q_b\!:\barbc_m{}^\dagger\rvac
=\Bigg( \tbb_m{}^\dagger
  -{1\over\sqrt{2}}\tbb_0\bc_0\barbc_m{}^\dagger
  +{m\over\sqrt{\pi}}\bc_0\barbc_m{}^\dagger\Bigg)\rvac
 \qquad (m>0), 
\end{eqnarray}
and hence
\begin{eqnarray}
\hspace*{-30pt}
\vwightman{\barbc_n (:\!\skew3\hat Q_b\!:)^2\barbc_m{}^\dagger}
&=&\vwightman{\tbb_n\tbb_m{}^\dagger}
 -\vwightman{\tbb_n\bc_0\Bigg( {1\over\sqrt{2}}\tbb_0
 -{m\over\sqrt{\pi}}\Bigg)\barbc_m{}^\dagger} \nonumber \\
&&
-\vwightman{\barbc_n\Bigg({1\over\sqrt{2}}\tbb_0
 -{n\over\sqrt{\pi}}\Bigg)\bc_0\tbb_m{}^\dagger} \qquad (n,\ m>0).
\end{eqnarray}
With the aid of \eqno(4,29), \eqno(4,30) and \eqno(6,13), we calculate 
each term of \eqno(6,15), and then find
\begin{eqnarray}
\vwightman{\barbc_n (:\!\skew3\hat Q_b\!:)^2\barbc_m{}^\dagger}
&=&{2\sqrt{2}n\over\sqrt{\pi}}\delta_{n\,m}\vwightman{\tbb_0} 
  \nonumber \\
&& + {2n\over\sqrt{\pi}}\delta_{n\,m}\vwightman{\tbb_0\barbc_0\bc_0} 
   + {2n\over\sqrt{\pi}}\delta_{n\,m}\vwightman{\tbb_0\bc_0\barbc_0} 
\nonumber \\
&=&0.
\end{eqnarray}
This complete the proof of \eqno(6,11).
\vskip50pt
\Sec{Unified treatment of infinite and finite strings}
In this section, we present the calculation of $Q_b{}^2$ and 
$\skew3\hat Q_b{}^2$ in the $x$-space.\foot(h,{Since our calculation is 
made under the generalized normal-product rule, we need not take normal 
products for $Q_b$ and $\skew3\hat Q_b$ explicitly.})
This approach enables us to calculate both infinite and finite strings
simultaneously.  We calculate
\begin{eqnarray}
&&A\equiv\vwightman{j_b{}^-(x_1)j_b{}^-(x_2)\barc^+(x_3)\barc^+(x_4)},\\
&&B\equiv\vwightman{\hj_b{}^-(x_1)\hj_b{}^-(x_2)\barc^+(x_3)
 \barc^+(x_4)}.
\end{eqnarray}
For simplicity, we write
\begin{eqnarray}
\wightman{1,\,2}&=&
\left\{
\begin{array}{lcl}
\displaystyle{ {i\over2\pi}{1\over x_1{}^+-x_2{}^+-i0} }
&\qquad 
&\hbox{for infinite string,} \\
\noalign{\vskip5pt}
\displaystyle{ 
{i\over2\sqrt{2}\,\pi}
 \cot\Bigg({x_1{}^+-x_2{}^+\over\sqrt{2}}-i0\Bigg) }
& 
&\hbox{for finite string.}
\end{array}\right.
\end{eqnarray}
Then the following identity hold:
\begin{eqnarray}
&&\wightman{1,\,2}\partial_+{}^{x_1}\partial_+{}^{x_2}\wightman{1,\,2}
 =2\partial_+{}^{x_1}\wightman{1,\,2}\cdot\partial_+{}^{x_2}
   \wightman{1,\,2}
  +{\gamma\over 2\pi i}\partial_+{}^{x_1}\wightman{1,\,2},\\
&&\partial_+{}^{x_1}\wightman{1,\,2}\cdot\partial_+{}^{x_2}
   \wightman{1,\,2}
 ={i\over 12\pi}[(\partial_+{}^{x_1})^3\wightman{1,\,2}
  +2\gamma\partial_+{}^{x_1}\wightman{1,\,2}], 
\end{eqnarray}
where
\begin{eqnarray}
\gamma&=&
\left\{
\begin{array}{lcl}
0 &\qquad &\hbox{for infinite string,} \\
\noalign{\vskip5pt}
1 &\qquad &\hbox{for finite string.}
\end{array}\right.
\end{eqnarray}
\par
Substituting \eqno(2,31) into \eqno(7,1), we have
\begin{eqnarray}
A&=&
-\vwightman{\barc^+(x_1)c^+(x_1)\partial_+c^+(x_1)\cdot
\barc^+(x_2)c^+(x_2)\partial_+c^+(x_2)\cdot\barc^+(x_3)\barc^+(x_4)} 
 \nonumber \\
&&+\vwightman{c^+(x_1)\partial_+\phi_M(x_1)\partial_+\phi^M(x_1)\cdot
c^+(x_2)\partial_+\phi_N(x_2)\partial_+\phi^N(x_2)\cdot
\barc^+(x_3)\barc^+(x_4)}.\nonumber\\
&&
\end{eqnarray}
Since all fields involved in \eqno(7,7) are free fields, it is 
expressible in terms of the 2-point functions
\begin{eqnarray}
&&\vwightman{c^+(x_1)\barc^+(x_2)}=\vwightman{\barc^+(x_1)c^+(x_2)}
 =\wightman{1,\,2}, \\
&&\vwightman{\partial_+\phi_M(x_1)\partial_+\phi^N(x_2)}
 ={i\over\,2\,}\delta_M{}^N\partial_+{}^{x_1}\wightman{1,\,2}
\end{eqnarray}
only.   We find
\begin{eqnarray}
A&=&
\bigg[
{1\over\,2\,}(D-2)\partial_1\wightman{1,\,2}\cdot
  \partial_2\wightman{1,\,2}
 \cdot\wightman{1,\,4}\wightman{2,\,3} 
 +\partial_1\wightman{1,\,2}\cdot\wightman{1,\,2}
  \wightman{1,\,4}\partial_2\wightman{2,\,3}\nonumber \\
&& \hspace*{10pt}
 +\wightman{1,\,2}\partial_2\wightman{1,\,2}\cdot
   \partial_1\wightman{1,4}\cdot\wightman{2,\,3} 
 -\wightman{1,\,2}^2\partial_1\wightman{1,\,4}\cdot
  \partial_2\wightman{2,\,3}\bigg]
\nonumber \\
&& 
 - \ (3\,\leftrightarrow\,4),
\end{eqnarray}
where $\partial_1\equiv\partial_+{}^{x_1}$.  
After some manipulation, $A$ can be rewritten as
\begin{eqnarray}
A&=&
\bigg[{1\over\,2\,}(D-10)\partial_1\wightman{1,\,2}\cdot
 \partial_2\wightman{1,\,2}\cdot\wightman{1,\,4}\wightman{2,\,3} 
  \nonumber \\
&& \hspace*{10pt}
 -4\wightman{1,\,2}\partial_1\partial_2\wightman{1,\,2}\cdot
   \wightman{1,\,4}\wightman{2,\,3}\bigg] 
 - \ (3\,\leftrightarrow\,4) \nonumber\\
&&  + \ \Delta,
\end{eqnarray}
where $\Delta$ is a total-divergence part given by
\begin{eqnarray}
\Delta&\equiv&
\bigg\{ 2\partial_1\partial_2
 [ \wightman{1,\,2}^2\wightman{1,\,4}\wightman{2,\,3} ] \nonumber \\
&& \hspace*{10pt}
-{3\over\,2\,}\partial_1[\wightman{1,\,2}^2\wightman{1,\,4}
  \partial_2\wightman{2,\,3}]
-{3\over\,2\,}\partial_2[\wightman{1,\,2}^2\partial_1\wightman{1,\,4}
  \cdot\wightman{2,\,3}]\bigg\} 
\nonumber \\
&& - \ (3\,\leftrightarrow\,4).
\end{eqnarray}
Then, by the help of \eqno(7,4) and \eqno(7,5), \eqno(7,1) finally becomes
\begin{eqnarray}
&&\hspace*{-20pt}
A=
\bigg\{
{i\over 24\pi}(D-26)(\partial_1)^3\wightman{1,\,2}
+\gamma{i\over12\pi}(D-2)\partial_1\wightman{1,\,2}\bigg\}
[\wightman{1,\,4}\wightman{2,\,3}-\wightman{1,\,3}\wightman{2,\,4}] 
 \nonumber\\
&& \  + \  \Delta.
\end{eqnarray}
\par
The calculation of \eqno(7,2) is straightforward but more lengthly 
because we encounter not only 2-point functions but also 3-point and 
4-point functions. We omit the details of the calculation.  
The result is simply
\begin{eqnarray}
&&B=\Delta.
\end{eqnarray}
\par
We return to \eqno(7,13).  First, we consider the infinite string.
From \eqno(2,32) ($j_b{}^+$ does not contribute) and \eqno(7,13), 
we have
\begin{eqnarray}
&&\hspace*{-50pt}
\vwightman{Q_b{}^2\barc^+(x_3)\barc^+(x_4)} \nonumber \\
&&\hspace*{-30pt} 
={i\over24\pi}(D-26){1\over\,2\,}\left({i\over2\pi}\right)^3
  \int_{-\infty}^{\infty}dx_1{}^1\int_{-\infty}^\infty dx_2{}^1 
 (\partial_+{}^{x_1})^3{1\over x_1{}^+-x_2{}^+-i0} \nonumber \\
&&
\bigg(
  {1\over x_1{}^+-x_4{}^+-i0}{1\over x_2{}^+-x_3{}^+-i0}
- {1\over x_1{}^+-x_3{}^+-i0}{1\over x_2{}^+-x_4{}^+-i0}
\bigg).
\end{eqnarray}
But \eqno(7,15) turns out to vanish, as is seen by carrying out the 
integration over $x_1{}^1$ as a contour integral in the lower 
half-plane.
This kind of reasoning applies to any Wightman function which has $Q_b$
at the left or right end.\foot(i,{Of course, sufficient damping of the 
integrand as $|x^1| \,\to\,\infty$ is needed.})
That is, we may infer that
\begin{eqnarray}
&&\lvac Q_b=0, \qquad Q_b\rvac=0
\end{eqnarray}
hold at the representation level.
\par
In order to have anomaly, therefore, we should consider 
$\vwightman{\barc^+(x_3)Q_b{}^2\barc^+(x_4)}$, for which $-i0$ is 
replaced by $+i0$ in all denominator factors involving $x_3{}^+$ in 
\eqno(7,15). We then find
\begin{eqnarray}
&&\vwightman{\barc^+(x_3)Q_b{}^2\barc^+(x_4)}
={D-26 \over 8\pi^2}{1\over (x_3{}^+-x_4{}^+-i0)^4}.
\end{eqnarray}
We note that this result can be reproduced also if we use \eqno(7,16) 
and \eqno(2,30):
\begin{eqnarray}
\vwightman{\barc^+(x_3)Q_b{}^2\barc^+(x_4)}
&=&\lvac\{\barc^+(x_3),\,Q_b\}\{Q_b,\,\barc^+(x_4)\}\rvac \nonumber \\
&=&\vwightman{\tcalT^+(x_3)\tcalT^+(x_4)} \nonumber \\
\noalign{\vskip5pt}
&=&{D-26\over 8\pi^2}{1\over (x_3{}^+-x_4{}^+-i0)^4}.
\end{eqnarray}
\par
Next, we consider the finite strings.  We calculate the integrations of 
\eqno(7,13) by 
substituting the formula [cf.\eqno(3,10)]
\begin{eqnarray}
&&\wightman{1,\,2}=-{1\over\sqrt{2}\,\pi}\Bigg[ {1\over\,2\,}
 +\sum_{n=1}^\infty e^{-in\sqrt{2}(x_1{}^+-x_2{}^+)}\Bigg].
\end{eqnarray}
For the closed string, the integration ranges of $x_1{}^1$ and $x_2{}^1$
are $[0,\;2\pi]$, while, for the open string, those are $[0,\;\pi]$ but
$j_b{}^+$ also contributes.  We obtain the same result for both strings.
\par
We can again infer \eqno(7,16) by using the fact that the summation 
over $n$ in \eqno(7,19) is restricted to $n>0$ because the integral
\begin{eqnarray}
&&\int_0^\pi dx_1{}^1\,\cos(nx_1{}^1 + mx_1{}^1)=\pi \delta_{n+m,\,0}
\end{eqnarray}
vanishes for $n,\ m>0$, as long as the purely zero-mode term of $Q_b$
does not contribute.  
A simple manipulation yields
\begin{eqnarray}
&&\hspace*{-20pt}
\vwightman{\barc^+(x_3)Q_b{}^2\barc^+(x_4)} 
={2\over\pi^2}\sum_{n=1}^\infty 
  \bigg[{D-26\over 24}n^3 -{D-2\over 24}n\bigg]
  e^{-in\sqrt{2}(x_3{}^+-x_4{}^+)}.
\end{eqnarray}
If we calculate $\vwightman{\barc^+(x_3)Q_B{}^2\barc^+(x_4)}$ by using
\eqno(6,1) with $\alpha_0=0$, we find that it precisely equals the 
right-hand side of \eqno(7,21).  Thus our approach is consistent with
Kato-Ogawa's one apart from the introduction of $\alpha_0$.
\par
Carrying out the summation over $n$ in \eqno(7,21), we obtain
\begin{eqnarray}
&&\vwightman{\barc^+(x_3)Q_b{}^2\barc^+(x_4)} \nonumber \\
&&\qquad
={D-26\over 32\pi^2}
  \Bigg(\sin{x_3{}^+-x_4{}^+\over\sqrt{2}}\Bigg)^{\!\!-4}
 +{1\over2\pi^2}
  \Bigg(\sin{x_3{}^+-x_4{}^+\over\sqrt{2}}\Bigg)^{\!\!-2}.
\end{eqnarray}
This result is precisely equal to 
$\vwightman{\tcalT^+(x_3)\tcalT^+(x_4)}$, as is verified by direct 
calculation.
\par
Finally, we consider the case of $\skew3\hat Q_b$.  
By the same reasoning as that of \eqno(7,16), we see
\begin{eqnarray}
&&\lvac\skew3\hat Q_b=0,\qquad\skew3\hat Q_b\rvac=0.
\end{eqnarray}
From \eqno(7,14), we have
\begin{eqnarray}
&&\vwightman{\barc^+(x_3)\skew3\hat Q_b{}^2\barc^+(x_4)}=0.
\end{eqnarray}
Corresponding to \eqno(7,18), \eqno(7,24) can be reproduced also by
considering
\begin{eqnarray}
\vwightman{\barc^+(x_3)\skew3\hat Q_b{}^2\barc^+(x_4)}
&=&\vwightman{\{\barc^+(x_3),\,\skew3\hat Q_b\}
   \{\skew3\hat Q_b,\,\barc^+(x_4)\}} \nonumber \\
&=&\vwightman{\tb^+(x_3)\tb^+(x_4)}=0.
\end{eqnarray}
\vfill\eject
\Sec{Discussion}
In the present paper, we have clarified how the Kato-Ogawa string 
theory can be understood in terms of our approach to the 
conformal-gauge two-dimensional quantum gravity.  
Our way of constructing Wightman functions reproduces the formulation 
of the Kato-Ogawa theory except for the introduction of 
a regularization parameter $\alpha_0$.
\par
We have shown that Kato-Ogawa's claim, $Q_B{}^2\not=0$ for $D\not=26$, 
is not a result intrinsic to the BRS quantization of the string theory. 
It is possible to construct explicitly a BRS charge nilpotent for any 
value of $D$.  {\it The BRS invariance itself is not anomalous.\/}
What is anomalous is the B field equation, which is anomaly-free only 
at the critical dimension $D=26$.  It should be noted, however, that 
the absence of the field-equation anomaly at $D=26$ is {\it not\/} 
a general property; in the de Donder gauge,the field-equation anomaly 
does not disappear for any value of $D$, as was shown 
elsewhere.\cite{AN1}
Even in that case, the BRS invariance is not broken.
As was clarified already,\cite{AN5} the appearance of the critical
dimension $D=26$ itself is {\it not\/} an intrinsic result in the 
de Donder-gauge two-dimensional quantum gravity.
\par
Our next problem is to reformulate the no-ghost theorem. 
Since Kato and Ogawa eliminated the B field from the outset, their 
treatment of the no-ghost theorem is rather different from the original 
form of the Kugo-Ojima quartet mechanism.\cite{Kugo-Ojima}  
We think that the B field should be adopted as a member of the 
Kugo-Ojima quartet.  It will be possible to construct the physical 
subspace explicitly as the proper framework of the two-dimensional 
quantum gravity rather than as the string theory.
\vfill\eject

%
%
\end{document}